\documentclass[11pt,epsf,letterpaper]{article}%
\usepackage{setspace}
\usepackage{color}
\usepackage{amsfonts,amssymb,amsmath,amsthm}
\usepackage{verbatim}
\usepackage{graphicx}
\usepackage{amsmath}
\usepackage{amsfonts}
\usepackage{amssymb}
\usepackage{color}%
\providecommand{\U}[1]{\protect\rule{.1in}{.1in}}

\begin{document}

\title{Complex angular momenta approach for scattering problems in the presence of
both monopoles and short range potentials}
\author{Fabrizio Canfora$^{1}$\\$^{1}$\textit{Centro de Estudios Cient\'{\i}ficos (CECS), Casilla 1469,
Valdivia, Chile.}\\{\small canfora@cecs.cl}}
\maketitle

\begin{abstract}
It is analyzed the quantum mechanical scattering off a topological defect
(such as a Dirac monopole) \textit{as well as} a Yukawa-like potential(s)
representing the typical effects of strong interactions. This system, due to
the presence of a short-range potential, can be analyzed using the powerful
technique of the \textit{complex angular momenta} which, so far, has not been
employed in the presence of monopoles (nor of other topological solitons). Due
to the fact that spatial spherical symmetry is achieved only \textit{up to
internal rotations}, the partial wave expansion becomes very similar to the
\textit{Jacob-Wick helicity amplitudes} for particles with spin. However,
since the angular-momentum operator has an extra "internal" contribution,
\textit{fixed cuts} in the complex angular momentum plane appear.
Correspondingly, the background integral in the Regge formula does not
decrease for large values of $\left\vert \cos\theta\right\vert $ (namely,
large values of the Mandelstam variable $s$). Hence, the experimental
observation of this kind of behavior could be a direct signal of non-trivial
topological structures in strong interactions. The possible relations of these
results with the soft Pomeron are shortly analyzed.

\end{abstract}

\newpage

\section{Introduction}

One of the most important non-perturbative effects in Field Theory is the
appearance of topological solitons \cite{manton} \cite{balabook} \cite{Shnir}
\cite{rossiR}. Such classical configurations are topologically stable and, for
this reason, are believed to play a crucial role at a quantum level as well
(the most important arena being, of course, the problem of color confinement:
for a pedagogical review on the role of topological configurations in the
problem of confinement\ see \cite{greenconf}). Such objects possess a
conserved topological charge which prevents them from being deformed to the
trivial vacuum. A key characteristic of such non-trivial topological
structures is that, due to the non-vanishing topological charge, spatial
spherical symmetry must be realized in a rather subtle way. As it was firstly
realized in modern terms by Skyrme when introducing his famous model
\cite{skyrme} (but the same argument also applies when dealing with local
internal symmetries), the most obvious way to describe a spherical object
would be to take all the fields of the Lagrangian to depend only on the radial
coordinate. However, in this case the topological charge would vanish (as an
easy computation shows immediately). Thus, the most one can get is to have
the\textit{ fields depending on the angular coordinates as well but in such a
way that energy density is spherically symmetric}. This, actually, is the
definition of hedgehog ansatz: the hedgehog is symmetric under spatial
rotations but only \textit{up to an internal symmetry transformation} (namely,
the spatial rotation needs to be compensated by an internal symmetry
transformation in order to achieve the sought invariance). This quite simple
characteristic is behind the remarkable intuition of Skyrme that his hedgehog
(called Skyrmion), despite being constructed in a purely Bosonic theory,
should be actually quantized as a Fermion. The reason is that the generators
of the angular momentum acquire extra contributions from the internal symmetry
transformation so that the eigenvalues are not required to be integer anymore
and can be also half-integers.

The simplest and most famous example in gauge field theory which gives rise to
similar phenomena is the Dirac monopole \cite{Dirac}. In the case of the Dirac
monopole too one gets spherical symmetry up to an internal (gauge) symmetry
transformation. This fact has highly non-trivial consequences when one
analyzes the Schrodinger (as well as the Klein-Gordon and Dirac) equation(s)
in the electromagnetic field\footnote{Here and in the following sections the
sentence "t\textit{he Schrodinger (Dirac, Klein-Gordon) equation(s) in the
electromagnetic field of a Dirac monopole}" will always mean the Schrodinger
(Dirac, Klein-Gordon) equation(s) with the minimal coupling $\partial_{\mu
}\rightarrow\nabla_{\mu}=\partial_{\mu}-ieA_{\mu}$ to the gauge potential
$A_{\mu}$ of a Dirac monopole.} of a Dirac monopole. It was soon realized that
the angular momentum operator in the case of the Schrodinger equation in the
field of a Dirac monopole (which was studied for the first time in
\cite{Banderet} and \cite{Tamm}) receives an extra contribution and that such
modified angular momentum operator is responsible for the fact that a scalar
charged particle in the field of a Dirac monopole can behave as a Fermion (for
a very clear analysis see \cite{sphericalupto1}). Such non-trivial feature of
the total angular momentum of a monopole is responsible of remarkable effects
not only in Quantum Field Theory (QFT henceforth) as already emphasized but
also in general relativity (see, in particular, \cite{BH}).

A very far-reaching generalization of the Dirac monopole has been constructed
in non-Abelian gauge theory coupled to a Higgs field in the adjoint
representation of the gauge group by `t Hooft and Polyakov \cite{thooft}
\cite{polyakov}. Such configurations are in a sense a regularized version of
the Dirac monopole as, far from the monopole core, they look like Dirac
monopoles while having, at the same time, regular cores (for a detailed
review, see \cite{manton} \cite{Shnir}). Also non-Abelian monopoles are
hedgehog in the sense that the corresponding gauge fields depend both on the
radial coordinate and on the angles but in such a way that the topological
charge (which in this case represents the non-Abelian magnetic charge) is
non-vanishing and, at the same time, the energy-density is spherically
symmetric. For this reason, the generators of the angular momentum acquire
extra contributions from the generator of the gauge symmetry. Hence, once
again, the eigenvalues are not required to be integers anymore and can be also
half-integers. This fact is at the basis of the well-known `t
Hooft-Hasenfratz-Jackiw-Rebbi-Goldhaber phenomenon \cite{sphericalupto2}
\cite{sphericalupto3} \cite{sphericalupto3.5}\ which states that a scalar
field charged under the gauge group is transformed into a Fermion in the field
of a `t Hooft-Polyakov monopole. Moreover, this effect persists even if the
effects of General Relativity are turned on \cite{meronBH}. The `t
Hooft-Polyakov monopole is not known analytically in the realistic
case\footnote{Exact solutions for the non-Abelian monopoles can be found in
the BPS approximation \cite{BPS1} \cite{BPS2} in which the Higgs potential
vanishes but one keeps the non-trivial boundary conditions.} in which the
Higgs potential is non-vanishing. In this case, \textit{far from the core},
one expects Yukawa-like behavior of the Higgs fields together with the
massless behavior of the gauge field associated with the unbroken $U(1)$ gauge
symmetry (which, asymptotically, looks exactly like a Dirac monopole).
Consequently, a very interesting analysis is the study of scattering from
\textit{both} Dirac monopole \textit{and} Yukawa`s potentials.

The above considerations show how crucial are the modified generator of the
angular momentum operator in the analysis of the physical properties of
topologically non-trivial configurations.

The most obvious and natural way to probe (both theoretically and
experimentally) this kind of objects is of course through scattering
\cite{dealfaro} \cite{reggebook1}\ \cite{reggebook2}. Due to the interest in
this non-trivial configurations, the scattering of charged particles on
monopoles and dyons has been deeply analyzed: very important references in
this respect (after the pioneering works \cite{Banderet} and \cite{Tamm}) are
\cite{Wheeler} \cite{Schwinger} \cite{Boulware} \cite{BalachandranM}. One of
the most important technical results (which will be very useful in the
following) achieved in these references has been the proper definition of the
partial wave expansion in the presence of a monopole. However, what was not
considered in these references is the inclusion of a (superposition of)
Yukawa`s potential(s): as it is clear from the above considerations, such an
inclusion is very natural and welcome from the point of view of non-Abelian
monopoles. Moreover, from a QFT perspective (see, for instance, \cite{charap})
when dealing with strong interactions, it is very natural to expect the
presence of (superpositions of) Yukawa`s potentials. Thus, a very natural
problem to consider (which, to the best of author`s knowledge, has not been
analyzed before) is the quantum mechanical scattering \textit{both} from a
monopole \textit{and} (a superposition of) Yukawa`s potential(s) which, at the
very least, describes the scattering processes far from the core of a `t
Hooft-Polyakov monopole.

At a first glance, one could think that such a problem is technically much
more complex than the original analysis of \cite{Wheeler} \cite{Schwinger}
\cite{Boulware} \cite{BalachandranM} as one cannot get neither the wave
function nor the spectrum in closed forms. However, a huge benefit (which, in
the opinion of the present author, vastly exceeds the above mentioned
technical disadvantages) of the presence of Yukawa potentials is that very
powerful tools become available: the Sommerfeld-Watson transform and the Regge
theory of complex angular momenta \cite{regge1} \cite{regge2} (which cannot be
applied directly to the scattering from monopoles or dyons alone mainly
because of the long-range effects of the monopoles). The applications of such
techniques (which have been extended to quantum field theory in \cite{Gribov}
\cite{froissartP} \cite{UniBound1} \cite{gellmann} \cite{qft1} \cite{qft2};
detailed reviews are \cite{Gribook} and \cite{ReggeReview1}) in the context of
strong interactions together with the Mandelstam representation
\cite{mandrepr}\ have been extremely successful. Moreover, due to the
fundamental role of the angular momentum within the Regge approach, it is
clear that such an approach is very suitable when dealing with scattering from
topologically non-trivial objects which are characterized (as it has been
discussed above) by modified angular momentum operators. For instance, a very
interesting by-product of the present analysis is that it disclose in a very
clear way the similarities (which should be expected, from the intuitive point
of view, on the basis of \cite{sphericalupto1}) between the \textit{helicity
amplitudes} for particles with spin introduced by Jacob and Wick in their
pioneering paper \cite{jacobi0} and the partial-wave expansion in the presence
of both monopoles and short range potential. However, there is also a crucial
difference between the two partial wave amplitudes which could be interpreted
as a fingerprint of the presence of non-trivial topological structures in
gauge theories as it will be discussed in the next sections.

Besides the intrinsic theoretical interest of this analysis, the application
of the theory of complex angular momenta when topological solitons are present
could be relevant for a very well known open problem in the field of strong
interactions: the Pomeron \cite{pomeranchuk} (detailed reviews are
\cite{reviewpomeron1} \cite{pomeronbook1}). The key issue is that\ at low
transferred momentum $t$ observations show that the scattering amplitude does
not decrease with $s$ as one would expect on the basis of the Froissart-Martin
bound \cite{UniBound1} \cite{UniBound2} (a modern and interesting analysis can
be found in \cite{UniBound3}). It is worth emphasizing that the
Froissart-Martin bound can be derived from very general hypothesis such as
unitarity, analyticity as well as from the \textit{short range nature of
strong interactions}. The present analysis shows that the hypothesis of short
range interactions is rather subtle when topological defects are present.
Indeed, one of the typical contributions in the Schrodinger (as well as
Klein-Gordon) equation(s) when a monopole is present is that the centrifugal
barrier increases. This implies, in particular, that the effects of the
monopole \textit{are not short-range as one can feel them even from very far}.
There are many approach to the Pomeron based on QCD, the most powerful being
the BFKL equation \cite{BFKL1} \cite{BFKL2} \cite{BFKL3} (detailed review are
\cite{pomeronbook2}\ \cite{LipaBook}). However, the issue related to the low
$t$ behavior is still open. As it will be explained in the next sections, the
present analysis suggests that the inclusion of the non-trivial topological
structures of the Yang-Mills vacuum within the BFKL formalism could be an
important step to solve this issue.

As far as the increase of the centrifugal barrier induced by topological
solitons is concerned, it is worth emphasizing\footnote{I thank the anonymous
referee for this interesting remark.} that, in 2+1 dimensions, a similar
effect does occur as well. In particular, the Chern-Simons term describing the
interaction of a charge with the (2+1)-dimensional magnetic monopole (see, for
instance, the detailed review \cite{wilczek}) manifests itself by changing the
plane geometry felt by a free particle into a (2+1)-dimensional geometry with
a conical defect. Not surprisingly, at classical level a charged particle
moving into the electromagnetic field generated by a Dirac monopole in (3+1)
dimensions also moves along conical surfaces (see \cite{plyus} and, for a
detailed review, the first two chapters of \cite{Shnir}).

The paper is organized as follows: in the second section, there is a short
review of the usual Watson-Sommerfeld-Regge transform. In the third section,
the Schrodinger equation in the presence of both a monopole and a short range
together with the corresponding scattering amplitude are discussed. In the
fourth section, the Watson-Sommerfeld-Regge transformation in the presence of
monopoles and short range potential is analyzed. In the fifth section, the
limitations of the present approach are presented. In the sixth section, the
relativistic generalization of the results are introduced and the relations
with the BFKL equation are shortly emphasized. Finally, in the last section
some conclusions and perspectives are included.

\section{Review of Watson-Sommerfeld-Regge transform}

\qquad Here all the necessary ingredients needed to perform the
Watson-Sommerfeld-Regge transform in quantum mechanical scattering are shortly
reviewed with particular emphasis on the technical steps which are going to
change in the presence of topological solitons. Let us begin with the usual
expression for the partial wave expansion of the scattering amplitude:%
\begin{align}
f\left(  \theta,k\right)   &  =\frac{1}{2ki}\sum_{l=0}^{+\infty}\left(
2l+1\right)  P_{l}\left(  \cos\theta\right)  \left[  \exp\left(  2i\delta
_{l}\left(  k\right)  \right)  -1\right]  \ ,\ E=\frac{k^{2}}{2M}%
\ ,\label{partialwave1}\\
k^{2}\psi &  =\left[  -\frac{d^{2}}{dr^{2}}+\frac{\lambda^{2}-\frac{1}{4}%
}{r^{2}}+V\right]  \psi\ ,\ \ \ \ \label{partialwave2}\\
\lambda &  =l+\frac{1}{2}\ ,\label{partialwave2.5}\\
V\left(  r\right)   &  =\int_{M_{0}}^{\infty}\sigma\left(  \rho\right)
\frac{\exp\left[  -\rho r\right]  }{r}d\rho\ ,\ \ \ M_{0}>0\ \Rightarrow V\leq
c_{0}\frac{\exp\left(  -M_{0}r\right)  }{r}\ ,\ \ c_{0}\neq0\ .
\label{partialwave3}%
\end{align}
The above setting corresponds to the partial wave expansion for the
Schrodinger equation with a central potential of a scalar particle. As usual,
$\delta_{l}\left(  k\right)  $\ is the phase shift which can be defined in
terms of the Jost functions of the radial Schrodinger equation comparing the
asymptotic behaviors (both at $r\rightarrow0$ and $r\rightarrow\infty$) of the
solution of Eq. (\ref{partialwave2}) with and without potential. The angular
equation gives rise, obviously, to the the spherical harmonics (and to the
Legendre polynomials) which also appear in the partial-wave expansion.

The above defined partial-wave expansion satisfies the optical theorem (which
is an important consequence of unitarity):%
\begin{equation}
\operatorname{Im}f\left(  \theta=0\right)  =\frac{k}{4\pi}\sigma_{Total}\ ,
\label{optical}%
\end{equation}
where $\sigma_{Total}$ is the total cross section. In order to derive this
result the only necessary ingredients are the completeness of the Legendre
functions and the fact that $\delta_{l}\left(  k\right)  $ is real for real
values of $l$ and $k$.

The important link between the radial and the angular equations (which plays a
fundamental role in the Regge transform) is the (seemingly innocuous) relation
in Eq. (\ref{partialwave2.5}). In fact, as it will be shown in the next
section, it is precisely this relation which is modified in the presence of
monopoles (due to the non-trivial extra contributions which appear in the
angular momentum operator).

One can also get a partial-wave expansion without introducing the Schrodinger
equation but just exploiting the invariance of the $S$-matrix under spatial
rotation (see, for a clear and pedagogical analysis, chapter 6 of
\cite{reggebook1}) and then expanding over the corresponding eigenvectors. The
choice of the potential in Eq. (\ref{partialwave3}), based on \cite{charap},
is the one appearing in the pioneering papers \cite{regge1}\ and
\cite{regge2}. However, the hypothesis on the potential can be somehow relaxed
(see for instance \cite{dealfaro} \cite{reggebook1}) keeping, of course, the
short-range nature of the potential.

Using standard arguments \cite{dealfaro} \cite{reggebook1}\ \cite{reggebook2}
of scattering theory from (superposition of) Yukawa`s potential together with
the complex angular momentum technique introduced in \cite{regge1}
\cite{regge2} \cite{bottino}\ one can prove that there exist a unique
interpolating function\ which is meromorphic in $\lambda$ in the half-plane
$\operatorname{Re}\lambda>0$ which reduces to the usual phase shift for
integer $l$. Moreover, the analyticity (meromorphy) domain of the phase shift
has been also extended to the full complex $\lambda$-plane in
\cite{cutmandelstam} \cite{cutnewton} \cite{predazziregge}.

An important corollary of the results derived in \cite{cutmandelstam}
\cite{cutnewton} which will be useful in the present framework is that when
$\lambda$ is real $\delta\left(  \lambda,k\right)  $ is \textit{always} real
(namely, not only when $\operatorname{Re}\lambda>0$):%
\begin{equation}
\lambda\in%
\mathbb{R}
\Rightarrow\delta\left(  \lambda,k\right)  \in%
\mathbb{R}
\ . \label{reality}%
\end{equation}
Such a conclusion only depends on the radial Schrodinger equation (see, for
instance, the first two sections of \cite{cutmandelstam}).

The arguments which lead to the conclusion that, for short-range potentials,
there exist a unique interpolating function $\delta\left(  \lambda,k\right)  $
which reduces to the physical value for integers $l$ and which is meromorphic
in $\lambda$ in the full complex $\lambda$ plane \textit{depend exclusively on
the radial Schrodinger equation} (\ref{partialwave2}). The angular part (as
well as the appearance of the Legendre polynomials) only enters into the game
after one plugs the phase shift into the partial-wave expansion in Eq.
(\ref{partialwave1}). It is in this last step that the link between the radial
and angular part (expressed as a relation between $\lambda$ and $l$ which, in
the usual case, is the one in Eq. (\ref{partialwave2.5})) becomes very
important. In particular, this implies that whenever there is scattering
process in which the radial Schrodinger equation has the form in Eq.
(\ref{partialwave2}) with a potential of the form in Eq. (\ref{partialwave3}),
it is possible to define (following exactly the same arguments of
\cite{regge1}, \cite{regge2}, \cite{bottino}, \cite{cutmandelstam} and
\cite{cutnewton}) a meromorphic phase-shift $\delta\left(  \lambda,k\right)  $
which satisfies all the properties and the bounds described in the above
mentioned references and which represents the scattering effects of the
potential in Eq. (\ref{partialwave3}) with respect to the free-waves (which
are, of course, the solutions corresponding to $V=0$).

Particular examples of the many results which only depend on the radial
Schrodinger equation are the two very important inequalities (see
\cite{regge1}, \cite{regge2} and \cite{bottino})%
\begin{align}
\left\vert \left[  \exp\left(  i\delta\left(  \lambda,k\right)  \right)
-1\right]  \right\vert  &  \leq\sigma\left(  k\right)  \frac{\exp\left(
-\alpha\lambda\right)  }{\lambda^{1/2}}\ ,\label{ine1}\\
\underset{\left\vert \operatorname{Im}\lambda\right\vert \rightarrow
\infty,\ \operatorname{Re}\lambda\rightarrow const}{\lim}\left[  \exp\left(
i\delta\left(  \lambda,k\right)  \right)  -1\right]   &  =0\ , \label{ine2}%
\end{align}
where$\ \cosh\alpha=1+M_{0}^{2}/2k^{2}$. Such inequalities are very relevant
both for the Regge analytic continuation of the Watson-Sommerfeld transform
and for the physical interpretation of the phase-shift. An elegant proof of
Eq. (\ref{ine2}) can be found in \cite{calogeroI}. Both inequalities are
relevant to derive the Regge formula for the scattering amplitude.

Since Eq. (\ref{partialwave2.5}) relates $\lambda$ (which is the variable
entering in the analytic continuation of the phase shift) with $l$ (which is
the discrete label which enters into the definition of the partial wave
expansion in Eq. (\ref{partialwave1})) through a simple linear (in particular,
\textit{analytic}) expression, the (analytically continued) phase shift is
meromorphic in the same variable which defines the (analytic continuation of
the) partial wave expansion.

In particular, all the factors (namely, $\left(  2l+1\right)  $, $P_{l}\left(
\cos\theta\right)  $ and $f_{\lambda}\left(  k\right)  $) which appear in each
term of the partial wave expansion in Eq. (\ref{partialwave1}) can be suitably
extended to \textit{analytic functions of the same variable} $\lambda$.
Consequently, one can go (see \cite{dealfaro} \cite{reggebook1}%
\ \cite{reggebook2}) from the expression in Eq. (\ref{partialwave1}) into the
Sommerfeld-Watson expression%
\begin{equation}
f\left(  \theta,k\right)  =\frac{1}{2\pi k}%
{\displaystyle\int\limits_{C}}
\frac{\lambda d\lambda}{\cos\pi\lambda}P_{\lambda-1/2}\left(  -\cos
\theta\right)  \left[  S\left(  \lambda,k\right)  -1\right]  \ \ ,\ \ \lambda
=l+\frac{1}{2}\ , \label{partialwave4}%
\end{equation}
and, then, to the Regge expression (thanks to the bounds for the phase-shift
found in \cite{regge1}, \cite{regge2} and \cite{bottino})%
\begin{equation}
f\left(  \theta,k\right)  =\frac{1}{2\pi k}%
{\displaystyle\int\limits_{-\infty}^{+\infty}}
\frac{\lambda d\lambda}{\cos\left(  i\pi\lambda\right)  }P_{i\lambda
-1/2}\left(  -\cos\theta\right)  \left[  S\left(  i\lambda,k\right)
-1\right]  -\frac{i}{k}%
{\displaystyle\sum\limits_{j=1}^{N}}
R_{j}P_{\alpha_{j}}\left(  -\cos\theta\right)  \ . \label{partialwave5}%
\end{equation}
In Eq. (\ref{partialwave4}) the circuit $C$ rounds counterclockwise the zeros
of $\cos\pi l$.

The representation in Eq. (\ref{partialwave5}) is called \textit{Regge formula
for the scattering amplitude}. The circuit of the first term on the right hand
side of Eq. (\ref{partialwave5}) corresponds to the line $\operatorname{Re}%
\lambda=0$ (together with a semi-circle at infinity which, however, does not
contribute due to the inequality (\ref{ine1})).

The first term on the right hand side of Eq. (\ref{partialwave5}) is called
\textit{background integral} and it gives small contribution for large
$\cos\theta$ as it decreases as $\left\vert \cos\theta\right\vert ^{-1/2}%
$\ (due to the fact that the Legendre polynomial within the integral in Eq.
(\ref{partialwave5}) has index $i\lambda-1/2$ with $\lambda$ real). As it has
been already emphasized, in the usual case, it is possible to push further to
the left the integration path of the background integral \cite{cutmandelstam}
\cite{cutnewton} \cite{predazziregge}. This allows to make the contribution
from the background integral even smaller (as the label $i\lambda-1/2$ of the
Legendre polynomial would go into $i\lambda-K$ with $K>1/2$). However, as it
will be shown in the next sections, the \textit{presence of monopoles
represents an obstruction in pushing the background integral to the left}.

The second term on the right hand side of Eq. (\ref{partialwave5}) corresponds
to the contributions arising from the Regge poles (all the factors-but the
Legendre polynomial evaluated at the Regge poles-have been packed into $R_{j}$).

At last, three important asymptotic properties of the Legendre functions which
are needed to prove the above well-known results are:%
\begin{align}
&  P_{l}\left(  \cos\theta\right)  \underset{l\rightarrow\infty}{\sim}%
\frac{\sigma(E)}{l^{1/2}}\exp\left[  \left\vert \operatorname{Im}%
\theta\right\vert l\right]  \ \ ,\label{legendre1}\\
&  \left\vert \frac{P_{l}\left(  -\cos\theta\right)  }{\sin\pi l}\right\vert
\underset{\operatorname{Im}l\rightarrow\pm\infty}{\leq}\frac{b(\theta
)}{l^{1/2}}\exp\left[  -\left\vert \operatorname{Re}\theta\operatorname{Im}%
l\right\vert \right]  \ \ ,\label{legendre2}\\
&  P_{l}\left(  z\right)  \underset{\left\vert z\right\vert \rightarrow\infty
}{\sim}a(l)z^{l}\ , \label{legendre3}%
\end{align}
where the precise forms of the bounded functions $\sigma\left(  E\right)  $,
$b(\theta)$ and $a(l)$ are not relevant.

\section{Scattering from Monopoles and Yukawa potentials}

\qquad In the presence of a Dirac monopole the angular momentum operator is
the following (see for a detailed review \cite{Shnir}):%
\begin{equation}
\overrightarrow{J}=\overrightarrow{l}-\mu\frac{\widehat{r}+\widehat{z}}%
{1+\cos\theta}=J_{\theta}\widehat{e}_{\theta}+J_{\varphi}\widehat{e}_{\varphi
}+J_{r}\widehat{e}_{r}\ ,\ \ \mu=\frac{eg}{c\hbar}\ ,\label{angularm1}%
\end{equation}
where $\overrightarrow{l}$ is the standard orbital angular momentum, $e$ and
$g$ are the electric and magnetic charges, $\widehat{r}$\ and $\widehat{z}%
$\ are the unit vectors in the radial and $z$ directions respectively, $\mu$
is the quantized strength of the monopoles%
\[
\left\vert \mu\right\vert =\frac{1}{2}\ ,\ 1\ ,\ ...,
\]
while the $\widehat{e}_{k}$\ are the spherical unit vectors. Without loss of
generality, we can assume that $\mu>0$ as the analysis with $\mu<0$ is
similar. The components of the total angular momentum operator
$\overrightarrow{J}$\ read%
\begin{equation}
J_{\theta}=\frac{1}{\sin\theta}\left(  i\partial_{\varphi}+\mu\left(
1-\cos\theta\right)  \right)  \ ,\ \ J_{\varphi}=-i\partial_{\theta
}\ ,\ \ J_{r}=-\mu\ .\label{components}%
\end{equation}
Consequently, the centrifugal barrier in the Schrodinger equation is
determined by $\overrightarrow{J}^{2}$. The eigenvectors and eigenvalues of
$\overrightarrow{J}$ are the so-called \textit{rotation matrices} or
\textit{generalized spherical harmonics }(the conventions and normalizations
coincide with \cite{Schwinger}\ and \cite{BalachandranM}):%
\begin{align}
&  \overrightarrow{J}^{2}d_{\mu m}^{(l)}=sd_{\mu m}^{(l)}\ ,\ \ x=\cos
\theta\ ,\ \ s=l\left(  l+1\right)  \ ,\ \ l=n-\mu\ ,\ n\in%
\mathbb{N}
\ ,\ l\geq\mu\ ,\ -l\leq m\leq l\ ,\label{angular2}\\
&  d_{\mu m}^{(l)}\left(  \theta,\varphi\right)  =N_{\mu lm}\exp\left[
i\left(  \mu+m\right)  \varphi\right]  \left(  1-x\right)  ^{\frac{\left(
\mu+m\right)  }{2}}\left(  1+x\right)  ^{\frac{\left(  \mu-m\right)  }{2}%
}P_{l-m}^{\left(  \mu-m,\mu+m\right)  }\left(  x\right)  \label{angular3}\\
&  P_{n}^{\left(  \alpha,\beta\right)  }\left(  x\right)  =\frac{\left(
-1\right)  ^{n}}{2^{n}n!}\left(  1-x\right)  ^{-\alpha}\left(  1+x\right)
^{-\beta}\frac{d^{n}}{dx^{n}}\left[  \left(  1-x\right)  ^{\alpha+n}\left(
1+x\right)  ^{\beta+n}\right]  \ ,\ l=\mu,\mu+1,\mu+2,...,\label{angular4}%
\end{align}%
\begin{equation}
N_{\mu lm}=\left[  \frac{\left(  l-m\right)  !\left(  l+m\right)  !}{\left(
l-\mu\right)  !\left(  l+\mu\right)  !}\right]  ^{-1/2}%
\ ,\label{normalization}%
\end{equation}
where the $P_{n}^{\left(  \alpha,\beta\right)  }$\ are the Jacobi polynomials,
the $d_{\mu m}^{(l)}$\ are the generalized spherical harmonics or $d$-rotation
matrices (obviously, when $\mu=0$ the above expressions reduce to the usual
Legendre Polynomials and spherical harmonics). An important property of the
Jacobi polynomials is that they are entire functions of the three indices
$\alpha$, $\beta$ and $n$. The $P_{n}^{\left(  \alpha,\beta\right)  }$ form a
complete set for any fixed $\alpha$ and $\beta$. As it has been already
emphasized (the cleanest discussion is probably the one in
\cite{sphericalupto1}) the extra term in the above defined angular momentum
operator $\overrightarrow{J}$ (which obviously is directly related to the
gauge field of the Dirac monopole) is responsible for the fact that (depending
on the strength of the Dirac monopole) one scalar particles in the field of a
monopole can behave as Fermions. This consideration together with the
fundamental role of the angular momentum in the Regge formalism \cite{regge1}
\cite{regge2}\ strongly suggest that it may be a good idea to try to apply
Regge theory in the presence of topological solitons. It is also interesting
to note that in the cases of Skyrmions and of a `t Hooft-Polyakov monopoles,
the corresponding angular momentum operators get similar extra terms related
to the topological charges (see \cite{skyrme} and \cite{sphericalupto2}
\cite{sphericalupto3} \cite{sphericalupto3.5}). Moreover, far from the core of
a `t Hooft-Polyakov monopole (see section 6 of \cite{Boulware}) the scattering
problem of a scalar field charged under the $SU(2)$ gauge group in the field
of the non-Abelian monopole itself reduces precisely to the scattering problem
off a Dirac monopole. This observation (together with the essential question
about the role of the short-range interactions related to the Higgs field in
the realistic case in which the potential is non-vanishing) has been one of
the main motivations of the present work.

Three crucial properties of the Jacobi polynomials (see \cite{azimov} as well
as the appendix of \cite{omnesalessandrini}) which allow the
Watson-Sommerfeld-Regge transform in the monopole case (as it will be
described in the next sections) are the analogue of Eqs. (\ref{legendre1}),
(\ref{legendre2}) and (\ref{legendre3}) for the Legendre polynomials:%
\begin{align}
&  P_{l}^{\left(  \alpha,\beta\right)  }\underset{l\rightarrow\infty}{\sim
}\frac{\sigma_{\alpha\beta}(E)}{l^{1/2}}\exp\left[  \left\vert
\operatorname{Im}\theta\right\vert l\right]  \ \ ,\label{jacob1}\\
&  \left\vert \frac{P_{l}^{\left(  \alpha,\beta\right)  }}{\sin\pi
l}\right\vert \underset{\operatorname{Im}l\rightarrow\pm\infty}{\leq}%
\frac{b_{\alpha\beta}(\theta)}{l^{1/2}}\exp\left[  -\left\vert
\operatorname{Re}\theta\operatorname{Im}l\right\vert \right]
\ \ ,\label{jacob2}\\
&  P_{l}^{\left(  \alpha,\beta\right)  }\underset{\left\vert z\right\vert
\rightarrow\infty}{\sim}a_{\alpha\beta}(l)z^{l}\ , \label{jacob3}%
\end{align}
where the precise forms of the bounded functions $\sigma_{\alpha\beta}\left(
E\right)  $, $b_{\alpha\beta}(\theta)$ and $a_{\alpha\beta}(l)$ are not
relevant. Thus, for fixed values of the quantized monopoles strength $\mu$,
the Jacobi polynomials satisfy exactly the same asymptotic bounds as the usual
Legendre ones. Not so surprisingly, the above asymptotic behavior of the
deformed spherical harmonics are some of the key mathematical properties which
allow the extension of the Regge formalism of complex angular momenta to
helicity amplitudes \cite{jacobi0} for particles with spin (see \cite{jacobi1}
\cite{jacobi1.5} \cite{jacobi2} \cite{jacobi3}; a nice and elegant
group-theoretical formulation of the Regge formalism for helicity amplitude is
\cite{toller} \cite{tolleretal} \cite{tolleretal2}). What is more interesting
is that the above asymptotic properties of the generalized spherical harmonics
will be also crucial for the extension of the Regge formalism to the
scattering problem in the presence of both monopoles and short range
potentials. This, perhaps, could have been guessed on the basis of
\cite{sphericalupto1}. However, a crucial difference between the usual
helicity amplitudes and the scattering amplitudes in the presence of monopoles
and short range potential (a sort of fingerprint of topologically non-trivial
configurations) will be apparent when discussing the analytic continuation
\textit{a la Regge} of the scattering amplitude.

\subsection{The Schrodinger equation with monopoles and short range
potentials}

Thus, the Schrodinger equation in the electromagnetic field of a Dirac
monopole $\mathbf{A}$ of the form
\begin{equation}
\mathbf{A}=\frac{g}{r}\frac{1-\cos\theta}{\sin\theta}\widehat{e}_{\varphi
}\ \label{sm0}%
\end{equation}
($g$ being the magnetic charge) and a central potential reads%
\begin{equation}
-\frac{1}{2M}\left[  \left(  \overrightarrow{\nabla}\right)  ^{2}-V\left(
r\right)  \right]  \Psi\left(  \overrightarrow{r}\right)  =E\Psi\left(
\overrightarrow{r}\right)  \ ,\ \ E=\frac{k^{2}}{2M}\ , \label{sm00}%
\end{equation}
where, according to the minimal coupling rule, one has%
\[
\overrightarrow{\nabla}=\overrightarrow{\partial}-ie\overrightarrow{A}\ .
\]
In the above formula, $\overrightarrow{\partial}$\ is the flat spatial
gradient and $\overrightarrow{A}$\ is the gauge potential of the Dirac
monopole in Eq. (\ref{sm0}). It is worth to note that the most elegant and
mathematically sound procedure to define the gauge potential of a Dirac
monopole is the one based on the theory of fiber-bundle introduced by Wu and
Yang in \cite{wuyang} \cite{wuyang2}. One of the main advantages of such
formulation is that it avoids the use of singular gauge potentials (such as
the one in Eq. (\ref{sm0})). On the other hand, as far as the Schrodinger
equation is concerned, the local and the Wu-Yang approaches produce the same
result (as it can be seen by comparing Eq. (53), section 11 of \cite{wuyang2}
with Eqs. (\ref{zt}) and (\ref{radialM}) here below).

After standard manipulations (see the original papers \cite{Banderet} and
\cite{Tamm} as well as the nice review \cite{Shnir}), it is possible to bring
Eq. (\ref{sm00}) into the following form%
\begin{equation}
-\frac{1}{2M}\left[  \frac{1}{r^{2}}\partial_{r}\left(  r^{2}\partial
_{r}\ \right)  -\frac{\left(  \overrightarrow{J}\right)  ^{2}-\mu^{2}}{r^{2}%
}-V\left(  r\right)  \right]  \Psi\left(  \overrightarrow{r}\right)
=E\Psi\left(  \overrightarrow{r}\right)  \ ,\ \ E=\frac{k^{2}}{2M} \label{zt}%
\end{equation}
where the total angular momentum operator $\overrightarrow{J}$\ is defined in
Eq. (\ref{angularm1}) while the potential is a superposition of Yukawa
potentials as in the original analysis of \cite{regge1}, \cite{regge2} and
\cite{bottino}:%
\[
V\left(  r\right)  =\int_{M_{0}}^{\infty}\sigma\left(  \rho\right)  \frac
{\exp\left[  -\rho r\right]  }{r}d\rho\ ,\ \ \ M_{0}>0\ \Rightarrow V\leq
c_{0}\frac{\exp\left(  -M_{0}r\right)  }{r}\ ,\ c_{0}\neq0\ .
\]
The above equation is separable and, with standard methods (see, for instance,
\cite{Shnir}), one gets the following radial and angular equations by
separation of variables%
\begin{equation}
\Psi\left(  \overrightarrow{r}\right)  =\frac{\psi_{k\lambda}\left(  r\right)
}{r}d_{\mu m}^{(l)}\left(  \theta,\varphi\right)  \label{separation}%
\end{equation}
:%
\begin{align}
k^{2}\psi_{k\lambda}  &  =\left[  -\frac{d^{2}}{dr^{2}}+\frac{\lambda
^{2}-\frac{1}{4}}{r^{2}}+V\right]  \psi_{k\lambda}\ ,\ \ \ \label{radialM}\\
\lambda &  =\lambda\left(  l\right)  =\sqrt{\left(  l+\frac{1}{2}\right)
^{2}-\mu^{2}}\ ,\ \ l\geq\mu\Rightarrow\lambda>\mu\ ,\label{radialM0.5}\\
sd_{\mu m}^{(l)}  &  =\overrightarrow{J}^{2}d_{\mu m}^{(l)}\ ,
\label{angularM}%
\end{align}
where the generalized spherical harmonics $d_{\mu m}^{(l)}$ have been
discussed in Eqs. (\ref{angular2}), (\ref{angular3}) and (\ref{angular4}). It
is worth emphasizing that, due to the inequalities in Eqs. (\ref{angular2})
and (\ref{radialM0.5}), the centrifugal barrier in the monopole case never
vanishes. This is one of the consequences of the fact that the generators of
the angular momentum operator get an extra contribution from the topologically
non-trivial configurations.

T\textit{he Regge formalism discloses a crucial difference between the usual
cases and the present case of a scattering problem off a topologically
non-trivial scatterer}. Such a difference is related to the functional
relation which links the parameter $\lambda$ which appears in the radial
Schrodinger equation with the parameter $l$ which determines both the
eigenvalues of the angular part and the discrete label of the partial-wave
series. As it has been already emphasized, in the usual case such link between
the radial and angular part (namely, Eq. (\ref{partialwave2.5})) is expressed
by an analytic relation between $\lambda$ and $l$. In the present case, the
link between the radial and angular equations is instead represented by Eq.
(\ref{radialM0.5}) which is not anymore an analytic relation between $\lambda$
and $l$. The origin of such a "mismatch" between $\lambda$ and $l$ is the fact
that the topological defect is spatially spherically symmetric only up to
internal transformation and, consequently, the centrifugal barrier is modified
in such a way to generate a fixed cut in the complex angular momentum plane.
This simple observation allows to determine a peculiar fingerprint of a
scattering process off a topologically non-trivial object as it will be
discussed in the next subsections.

\subsection{Partial-wave expansion for scattering off monopoles and short
range potentials}

Here, the analysis of chapter 6 of \cite{reggebook1} will be followed taking
into account that, in the presence of a topological soliton such as a
monopole, the angular momentum gets an extra contribution from internal
symmetries. In the present case, the scattering matrix $S$ is invariant under
the following rotation operator $R(\alpha)$:%
\begin{align*}
R  &  =R\left(  \alpha\right)  =\exp\left(  -i\alpha\overrightarrow{J}\right)
\ ,\\
R^{\dag}SR  &  =S\ ,
\end{align*}
where $J$ is the total angular momentum defined in Eq. (\ref{angularm1})
(which receives contributions from the monopole as well). The next step in
order to define the partial-wave expansion is to define the \textit{free
waves} or \textit{asymptotic states}. The best choice is, obviously, to define
as free waves the eigenfunctions of $H_{0}$, $J^{2}$ and $J_{z}$. However, it
is worth emphasizing that this choice implies that one has to choose as
\textit{free Hamiltonian} $H_{0}$ the Hamiltonian without the potential
(namely, $V=0$ in Eqs. (\ref{zt}) and (\ref{radialM})) but with the
\textit{effects of the monopole} (which are encoded in the modified
centrifugal barrier and in the replacement of the Legendre polynomials with
the Jacobi polynomials) \textit{included}.

The reason why this choice is very convenient\footnote{This is actually
mandatory if one wants to apply Regge theory when there is both a topological
soliton and a short range potential: the effects of topological solitons are
not short-range as one needs in Regge theory. However, as they can be encoded
into the angular momentum, such effects can be included in the free
hamiltonian. Thus, the interaction term is now short range and the Regge
theory can be applied.} is that if one would include the effects of the
monopole into the interaction, then the angular symmetry generators of the
free Hamiltonian (which, in such a case, would be just the free Laplacian)
would be different from the angular symmetry generators of the full
Hamiltonian. A more concrete way to see this is to notice that the effects of
the monopole are not small as they are described by a parameter ($\mu$ in our
case) which is quantized, thus it cannot be reduced continuously to zero. Even
from the perspective of the Bethe-Salpeter equation with singular potentials
(see \cite{bastaitonin} \cite{bastaitonin2}), the present choice is mandatory.
Namely, in order to construct well-defined solutions for the Bethe-Salpeter
equation with potentials which at the origin behave as $1/r^{2}$ (which is
basically the present case; on the other hand the authors of that references
also dealt with more singular potential at the origin), one must include the
singular potential into the free Hamiltonian (see in particular section 3 of
\cite{bastaitonin}). Thus \cite{bastaitonin} \cite{bastaitonin2} tell that, in
the case of a monopole and a short range potential, the free Hamiltonian is
the one without the short range potential but with the effects of the monopole
included\footnote{It is worth to emphasize that to take as free Hamiltonian
the one with the effects of the monopole included is a mathematically
well-defined procedure as the chosen free Hamiltonian has not bound states. In
the cases in which bound states are present, some extra care is needed.}.

Moreover, the standard procedure (for a pedagogical review see
\cite{instanton}) to compute any observable in the path integral formalism is
to consider the perturbative expansion \textit{within the given topological
sector}. This means, in particular, that the definition of free wave functions
depend on the topological sector which one is considering. There is an
undeniable physical reason for this already in the case of a Dirac monopole.
One can feel the effects of the monopole even from arbitrary far (for a
pedagogical review see the first two chapters of \cite{Shnir}) both
classically (as a classical particle in the field of a monopole is constrained
to move on a cone) and quantum mechanically (for instance, there is no
$s-$wave in the field of the monopole as the centrifugal barrier never
vanishes). This is why the definition of "free wave function" should depend on
the topological sector one is considering.

Therefore, in the cases in which there is both a monopole and a short range
potential, it only makes sense to consider as "free waves" the wave functions
in the presence of the monopole but without the potential. In other words, in
the present case, the \textit{free waves} \textit{correspond to the scattering
monopole wave functions constructed in the foundational references}
\cite{Banderet} \cite{Tamm} \cite{Wheeler} \cite{Schwinger} \cite{Boulware}.

For the above reasons, the spherical basis which will be used to diagonalize
the $S$ matrix will be%
\begin{align}
\left\langle \overrightarrow{x}\right.  \left\vert E,l,m\right\rangle  &
=i^{j}\left(  \frac{2m}{\pi p}\right)  ^{1/2}\frac{\widehat{j}_{\lambda
}\left(  kr\right)  }{r}d_{\mu m}^{(l)}\ ,\label{sm1}\\
\left\langle E`,l`,m`\right.  \left\vert E,l,m\right\rangle  &  =\delta\left(
E`-E\right)  \delta_{ll`}\delta_{mm`}\ , \label{sm2}%
\end{align}
where $\widehat{j}_{\lambda}\left(  kr\right)  $\ are the
Riccati-Bessel\ function while the $d_{\mu m}^{(l)}$\ are the rotation
matrices in Eq. (\ref{angularM}). Hence, the phase shift and the scattering
amplitude constructed in the present paper describe the effects of the short
range potential on top of the monopole scattering wave functions of
\cite{Banderet} \cite{Tamm} \cite{Wheeler} \cite{Schwinger} \cite{Boulware}.
In this basis, the $S-$matrix is diagonal as it commutes with $H_{0}$ and
$\overrightarrow{J}$:%
\begin{align}
\left\langle E`,l`,m`\right\vert \ S\ \left\vert E,l,m\right\rangle  &
=\delta\left(  E`-E\right)  \delta_{ll`}\delta_{mm`}S_{l}\left(  E\right)
\ ,\label{sm3}\\
S_{l}\left(  E\right)   &  =\exp\left[  2i\delta\left(  l,E\right)  \right]
\ , \label{sm4}%
\end{align}
where in Eq. (\ref{sm4}) it has been explicitly taken into account the
unitarity of the $S-$matrix. From this definition, one can immediately derive
a partial wave expansion expressing the scattering amplitude in the momentum
basis (which is directly related to the observable cross section: see
\cite{reggebook1} for a pedagogical exposition). The only technical detail
which must be taken into account is that, as for the usual spherical
harmonics, also the deformed spherical harmonics form a complete set. Thus,
the partial wave reads in this case%
\begin{equation}
f_{\mu}\left(  k,x\right)  =\frac{\exp\left(  -i\pi\mu\right)  }{k}\left(
\frac{1-x}{2}\right)  ^{\mu}\sum_{l=0}\left[  2\left(  l+\mu\right)
+1\right]  \exp\left[  i\delta\left(  l,k\right)  \right]  \sin\delta\left(
l,k\right)  P_{l}^{2\mu,0}\left(  x\right)  \ . \label{sm5}%
\end{equation}
The optical theorem (which discloses one of the most relevant differences
between the usual case and the Regge theory in presence of topological
solitons) will be discussed in the next sections separately for the $\mu=1/2$
case and the higher values of $\mu$.

In the "Reggeization" of helicity amplitudes (see in particular
\cite{jacobi1.5} \cite{jacobi2} \cite{jacobi3}) it is customary to define the
scattering amplitude without the "kinematical factors" $\left(  \frac{1-x}%
{2}\right)  ^{\mu}$ of the rotation matrices which may lead to extra
singularities in the scattering amplitude like branch cuts. On the other hand,
if one only keeps the Jacobi polynomial in the definition of the scattering
amplitude then only dynamical singularities appear. In the following only
dynamical singularities will be discussed.

\subsection{Phase shift and Schrodinger equation}

The only missing piece of information that one needs to discuss the scattering
amplitude using the Regge transform is the phase shift defining the $S$-matrix
elements which enters in the scattering amplitude in Eqs. (\ref{sm4}) and
(\ref{sm5}).

The radial Schrodinger equation (\ref{radialM}) is the same as in the usual
case in Eq. (\ref{partialwave2}). Thus, using the same arguments of
\cite{bottino}\ (based on the analysis of the Jost functions associated to Eq.
(\ref{radialM})) one can define a unique interpolating phase shift
$\delta\left(  \lambda,k\right)  $ which is meromorphic in the full complex
$\lambda$-plane (following the same arguments in \cite{cutmandelstam}
\cite{cutnewton}) and which fulfills all the bounds derived in \cite{regge1}
\cite{regge2} \cite{bottino}. One can also prove that the number of Regge
poles is finite (in the case of Yukawa`s potential) and that the inequalities
in Eqs. (\ref{ine1}) and (\ref{ine2}) are satisfied. Consequently (taking into
account the well-known properties of the Jacobi polynomials in Eqs.
(\ref{jacob1}), (\ref{jacob2}) and (\ref{jacob3}) which are completely
analogous to the properties of the usual Legendre polynomials needed for the
Regge transform) \textit{all the mathematical ingredients necessary in order
to perform the Watson-Sommerfeld-Regge transform of the scattering amplitude
in Eq. (\ref{sm5}) are available}.

However, the big difference with respect to the usual case is that the
variable $\lambda$ is related to the label $l$ of the angular functions by a
non-analytic relation: Eq. (\ref{radialM0.5}). Hence, the phase-shift
$\delta\left(  l,k\right)  =\delta\left(  \lambda\left(  l\right)  ,k\right)
$ as function of the angular label $l$ is meromorphic in the complex
$l$-plane\textit{ with a branch cut on the real} $l$ axis from $-\left\vert
\mu\right\vert -1/2$ to $\left\vert \mu\right\vert -1/2$. Therefore, unlike
what happens in \cite{cutmandelstam} \cite{cutnewton}, it will not be possible
to push further to the left the integration path of the background integral in
the Regge formula with monopoles.

It is also worth emphasizing the difference of the present situation with
respect to previous analysis of the Regge formalism with singular potentials
at the origin (see, in particular, \cite{fixedcut1} \cite{fixedcut2}
\cite{fixedcut3} \cite{fixedcut4} \cite{fixedcut5}) in which the usual Regge
formalism (namely, without any topological soliton and keeping both the
standard generators of the angular momentum and the Legendre polynomial in the
partial wave expansion) was extended to include potentials of the form
\begin{equation}
V_{S}=\frac{A}{r^{2}}\ . \label{singpot}%
\end{equation}
The physical results and their interpretations within that references (see, in
particular \cite{fixedcut1} \cite{fixedcut2} \cite{fixedcut3}) are basically
opposite to the present results (which strongly depend on the role of the
monopole and of the corresponding modified angular momentum generator and
Jacobi polynomials). In particular, in the usual case with singular potential
considered in \cite{fixedcut1} \cite{fixedcut2} \cite{fixedcut3} one has to
interpret the singular potential as attractive when $A>0$ and as repulsive
when $A<0$. Consequently, in that references, the fixed cut in the angular
momentum plane\footnote{In Regge theory applied to QFT, the appearance of cuts
in the complex $\lambda$ plane is a well known phenomenon (see, in particular,
the discussion in \cite{cut1} \cite{cut2} \cite{cut3}). The usual situation is
that such branch points are \textit{moving} (in the sense that they depend on
the Mandelstam variables). On the other hand, the cuts considered here are
fixed due to their topological origin.} which is generated by the singular
potential is interpreted as a sign of the attractive nature of the singular
potential itself. A further consequence of the results in that references is
that one should not consider the case $A>1/4$ as "the system collapse into the
center and the very concept of scattering is no longer clear" (this quotation
is from \cite{fixedcut3} but similar results are contained in \cite{fixedcut1}
\cite{fixedcut2} \cite{fixedcut5}).

On the other hand, the present situation is practically the opposite. At a
first glance, if one look at the Schrodinger equation in the presence of both
a monopole and a short range potential in Eq. (\ref{zt}) one could naively
think that the effect of the monopole is to decrease the centrifugal barrier
due to the extra potential%
\begin{equation}
V_{M}=\frac{\mu^{2}}{r^{2}} \label{singpot1}%
\end{equation}
which has always the opposite sign with respect to the usual centrifugal
barrier. In fact, as it is well known, this interpretation is completely wrong
since the effect of the monopole (as well as of all the relevant topological
solitons in 3+1 dimensions) is actually to increase the centrifugal barrier.
The information about the increase of the centrifugal barrier lies in the link
between the radial and angular equations. Unlike the setting considered in
\cite{fixedcut1} \cite{fixedcut2} \cite{fixedcut3} \cite{fixedcut5}, the
effects of the monopole \textit{do not reduce just to the extra term} in Eq.
(\ref{singpot1}). The angular equation and the corresponding eigenvalues are
affected by the presence of the monopole as well. This fact has important
consequences: first of all, in the present case there are not extra bound
states so that the scattering problem is well defined for any physical value
of $\mu$. Secondly, as far as the cut is concerned, due the fact that $\mu
\geq1/2$ we are precisely in the range which was not considered in
\cite{fixedcut1} \cite{fixedcut2} \cite{fixedcut3} \cite{fixedcut5} (namely,
$A>1/4$ in the notation of \cite{fixedcut3}) and yet, of course, there will be
no "collapse of the wave function on the center" (as guaranteed by the angular
part of the monopole-short range potential problem).

The physical consequences of these facts will be discussed in the next section.

\section{Monopole Regge formula and the fixed cut}

In the previous section it has been shown that one can define the phase shift
for a scattering problem off a monopole and a short range potential and that
the phase shift satisfies, in the natural variable $\lambda$, all the bounds
derived in the classic papers \cite{regge1} \cite{regge2} \cite{bottino}
\cite{cutmandelstam} \cite{cutnewton}. Such results together with the well
known bounds on the Jacobi polynomial (see Eqs. (\ref{jacob1}), (\ref{jacob2})
and (\ref{jacob3}): useful references are \cite{azimov} as well as the
appendix of \cite{omnesalessandrini}) do allow a Watson-Sommerfeld-Regge
transform. However, in the present case, there is an extra ingredient (which
is a \textit{characteristic fingerprint of the presence of topological
solitons in a scattering problem}) which is absent in the usual helicity
amplitudes analyzed in \cite{jacobi1} \cite{jacobi1.5} \cite{jacobi2}
\cite{jacobi3}.

Due to the fact that the relation between $\lambda$ and $l$ has the form in
Eq. (\ref{radialM0.5}), in the monopole case there is a fixed cut in the
complex $l$ plane along the real $l$ axis from $-\left\vert \mu\right\vert
-1/2$ to $\left\vert \mu\right\vert -1/2$. Thus, \textit{such a fixed cut
opens up due to the presence of a monopole and its size is} $2\left\vert
\mu\right\vert $. Consequently, as $2\left\vert \mu\right\vert $\ is at least
$1$, the cut will touch at least one of the poles of the factor $1/\cos
\pi\lambda$ which appears in the Sommerfeld-Watson transform (see Eq.
(\ref{partialwave4})). This fact prevents one from pushing the integration
path of the background integral in the Regge transform further to the left.
Thus, the procedure of \cite{jacobi3}\ (which is common for helicity
amplitudes) cannot be used in this case. In a sense, the presence of monopoles
can be seen as an obstruction to the maximal analyticity principle in the
complex $l$ plane advocated by Chew and Frautschi \cite{maximalanal1}
\cite{maximalanal2}.

It is worth emphasizing that the results of \cite{BalachandranM} (in which the
authors only analyzed the scattering off a monopole without any short-range
interaction and, consequently, without using the Regge approach) are closely
related to the present ones. The reason is that the origin of the branch cut
singularities of the scattering amplitude in $\cos\theta$ found in
\cite{BalachandranM} is the same as the origin of the fixed branch cut in the
complex $l$ plane found in the present work: both of them are related to the
mismatch in Eq. (\ref{radialM0.5}) between $\lambda$ and $l$ in the presence
of a non-trivial topological structure. It seems that the Regge theory is a
more suitable tool to disclose the physical implication of such a result (as,
for instance, it sheds considerable light on the asymptotic behavior of the
scattering amplitude).

\subsection{The case $\mu=1/2$}

Let us first consider the case in which $\mu=1/2$ (which is the lowest
possible non-trivial value of the monopole strength). Besides the intrinsic
interest of this case, from the point of view of non-Abelian theories, it
describes the asymptotic behavior of a scattering off a 't Hooft-Polyakov
monopole of unit charge (namely, the stable non-Abelian magnetic monopole) far
from its core \cite{Boulware}.

The first consequence of the presence of the fixed cut along the real $l$ axis
is that, in order to perform the Sommerfeld-Watson transform the analogue of
the circuit $C$ in Eq. (\ref{partialwave4}) (which in the usual case rounds
all the integers including $0$) now must exclude $0$ (namely, the first term
in the partial wave expansion in Eq. (\ref{sm5})) as the cut begins at $l=-1$
and ends at $l=0$.

Nevertheless, it is important to emphasize (as one can check directly) that
due to both the unitarity of the $S-$matrix elements (namely $\delta\left(
\lambda,k\right)  $ is real for $k$ and $\lambda$ real\footnote{It is useful
to remind that the validity of such a result only depends on the radial
Schrodinger equation (see, for instance, the first two sections of
\cite{cutmandelstam}). Consequently, it also holds in the present case.}; in
particular, for $k$ real, $\delta\left(  \lambda(l),k\right)  $ is real for
$l=0$, $1$, $2$,... so that in each term of the partial wave expansion in Eq.
(\ref{sm5}) the phase shift is real, see Eq. (\ref{reality})) and to the
completeness relations satisfied by the Jacobi polynomial the above scattering
amplitude satisfies the optical theorem:%
\[
\operatorname{Im}f\left(  k,\theta=0\right)  =\frac{k}{4\pi}\sigma_{Total}\ .
\]

Thus, the Sommerfeld-Watson transform of Eq. (\ref{sm5}) (in this section the
extra kinematical factors of the rotation matrices will not be considered)
reads%
\begin{align}
\left.  f_{\mu}\left(  k,x\right)  \right\vert _{\mu=1/2}  &  =2\exp\left[
i\delta\left(  0,k\right)  \right]  \sin\delta\left(  0,k\right)  P_{0}%
^{1,0}\left(  x\right)  +\nonumber\\
&
{\displaystyle\int\limits_{C_{1/2}}}
\frac{\left[  2\left(  l+\mu\right)  +1\right]  dl}{\sin\pi l}P_{l}%
^{1,0}\left(  -\cos\theta\right)  \left[  S\left(  \lambda\left(  l\right)
,k\right)  -1\right]  \ , \label{SomWatM1}%
\end{align}
$\ $where it has been taken into account that $\lambda\left(  l=0,\mu=\frac
{1}{2}\right)  =0$ and it has been explicitly emphasized that the phase shift
depends on $l$ through $\lambda$. Thus, the first term of the partial wave
expansion cannot be encoded in the Sommerfeld-Watson integral. On the other
hand, such a term is well behaved as the phase shift $\delta\left(
0,k\right)  $ is well behaved (and approaches to $0$) for large $k$ (see, for
instance, \cite{dealfaro}).

The Sommerfeld-Watson integral in Eq. (\ref{SomWatM1}) can be written in the
Regge form (\cite{regge1} \cite{regge2} \cite{bottino}):%
\begin{equation}
\left.  f\left(  k,x\right)  \right\vert _{\mu=1/2}=2\exp\left[
i\delta\left(  0,k\right)  \right]  \sin\delta\left(  0,k\right)  P_{0}%
^{1,0}\left(  x\right)  + \label{SomWat2}%
\end{equation}%
\[
+\frac{1}{2\pi k}%
{\displaystyle\int\limits_{C_{\varepsilon}}}
\frac{\left[  2\left(  l+\mu\right)  +1\right]  dl}{\sin\pi l}P_{l}%
^{1,0}\left(  -x\right)  \left[  S\left(  \lambda\left(  l\right)  ,k\right)
-1\right]  -\frac{i}{k}%
{\displaystyle\sum\limits_{j=1}^{N}}
R_{j}P_{\alpha_{j}}^{2\mu,0}\left(  -x\right)  \ ,\ \varepsilon\in%
\mathbb{R}
_{+}.
\]
In the above equations, the analogue of the $\operatorname{Re}\lambda=0$ line
for the background integral in the usual case (see Eq. (\ref{partialwave5}))
is the line $C_{\varepsilon}$ defined by the equation
\[
\operatorname{Re}l=\varepsilon\ ,
\]
where $\varepsilon$ is a positive arbitrarily small but non-vanishing number
(one cannot take $\varepsilon=0$ otherwise one would touch the cut which
begins at $l=0$). Moreover, it is easy to check that when $\operatorname{Im}%
l\rightarrow\pm\infty$ with $\operatorname{Re}l=\varepsilon$, one gets
\begin{equation}
\lim\left(  \operatorname{Im}\lambda\right)  \rightarrow\pm\infty
\ ,\ \lim\left(  \operatorname{Re}\lambda\right)  \rightarrow\frac{1}%
{2}\ ,\ \label{usefulsignature}%
\end{equation}
so that the well known bounds (see \cite{calogeroI} \cite{dealfaro}) for
$\delta\left(  \lambda,k\right)  $ when $\left\vert \operatorname{Im}%
\lambda\right\vert \rightarrow\infty$ with $\operatorname{Re}\lambda$ fixed
(which ensure the convergence of the background integral) can be applied

Thus, the remarkable effect is that, unlike what happens in the usual cases
\cite{regge1} \cite{regge2}, the background integral does not decrease anymore
for large values of $\left\vert \cos\theta\right\vert $. The obvious reason is
that the Jacobi polynomial in the background integral in Eq. (\ref{SomWat2})
is $P_{\varepsilon+iy}^{1,0}\left(  -x\right)  $ (with $\varepsilon$
arbitrarily small, real and positive and $y$ real varying from $-\infty$ to
$+\infty$). Hence, one does not get the usual decreasing behavior for large
values of $\left\vert \cos\theta\right\vert $. Rather, an oscillating
non-decreasing behavior for large values of $\left\vert \cos\theta\right\vert
$ appears. Thus, at low transferred momentum the behavior of the scattering
amplitude is not dominated by the leading Regge trajectory and the background
integral also plays a key role (compared with the standard case). It is only
when the leading Regge trajectory $\alpha_{P}\left(  t\right)  $ (which, in a
relativistic context, would correspond to the Pomeron) becomes positive enough
that the Regge pole behavior dominates the background integral.

\subsection{Monopoles of higher charges}

Here the case of higher values of $\mu$ are analyzed. From the point of view
of non-Abelian theories, this would correspond to (the large $r$ behavior of)
non-Abelian monopoles which are spherically symmetric and, at the same time,
have higher topological charges. In fact, such spherically symmetric hedgehog
configurations do not exist\footnote{Stable multi-monopoles solutions with
higher topological charges do indeed exist. However, such solutions are not
spherically symmetric (see \cite{Shnir}). Consequently, when stable
non-Abelian multi-monopole configurations are involved, the partial wave
expansion should be modified in order to take into account the lack of
spherical symmetry.}. An easy way to argue that this is the case is to observe
that such configurations would be highly unstable (the same is true even if
one gives up spherical symmetry requiring just axial symmetry: see
\cite{Shnir} \cite{rossiR}).

As from the point of view of non-Abelian theories the case of $\mu>1/2$ is
completely different from the case $\mu=1/2$, it is natural to wonder whether
the present approach is able to detect in a direct way such a difference.
Interestingly enough, within the present approach, the difference between
these two cases emerges in a very natural and intuitive way. When
$\mu=1,3/2,...$ it is trivial to see that the first terms (corresponding to
the lowest $l$) in the partial wave expansion in Eq. (\ref{sm5}) would have a
complex $\lambda\left(  l\right)  $ and, consequently, a complex phase shift
$\delta(\lambda(l),k)$ which signals instabilities (see the nice discussion in
\cite{beltrametti}). In particular, the optical theorem does not hold in such
cases. Thus, one can derive a well defined partial wave expansion and Regge
continuation to complex angular momenta only in the presence of stable
topological defects.

\section{On the range of validity of the approach}

As discussed in details in \cite{Boulware}, the scattering from a Dirac
monopole provides with a accurate description of the scattering form
non-Abelian monopole far from the core of the latter. In terms of energy
bounds, this means that the present results can be a valid description of
scattering process in non-Abelian theories when%
\begin{equation}
s<M_{mon}\ , \label{cond1}%
\end{equation}
where $s$ is the center-of-mass Mandelstam variable and $M_{mon}$ is the mass
scale which characterizes the non-trivial topological structure of the
non-Abelian theory.

The second limitation on the range of applicability of the present results in
non-Abelian theories arises from the requirement that loops effects should be
small. A necessary condition in order for this to happen in two-body
scattering is that%
\begin{equation}
\left\vert \frac{t}{s}\right\vert \ll1\ , \label{cond2}%
\end{equation}
where $t$ is the transferred momentum Mandelstam variable. It is worth to
emphasize that the above condition is nothing but the definition of the Regge region.

It is interesting to note that one of the most interesting and still open
issues in the application of Regge theory in QCD lies well within the range
defined by Eqs. (\ref{cond1}) and (\ref{cond2}). Such an issue has to do with
the soft Pomeron \cite{pomeranchuk} (detailed reviews are
\cite{reviewpomeron1} \cite{pomeronbook1}). Consequently, the present results
on the effects of topological solitons in Regge theory (which can be applied
in QFT only when the condition (\ref{cond1}) holds and $t$ is small enough to
make negligible all loops corrections) can be relevant from the soft Pomeron perspective.

It is by now quite clear that the actual behavior of the scattering amplitude
at $t=0$ is much more complicated than the one in \cite{Landsh1}\ (see, for
instance, \cite{refinedPomeron} \cite{refinedpomeron2}). Even if it is
well-known (see \cite{UniBound3} and references therein) that, at the scale of
the soft Pomeron, the total cross section is actually very far from saturating
the unitarity bound \cite{UniBound1} \cite{UniBound2} there is still a puzzle
related to the fact that while other trajectories lead to falling cross
sections, the Pomeron can lead to rising cross sections. In particular, it is
difficult to justify this kind of behavior in terms of a simple Regge pole.
Moreover, the BFKL equation \cite{BFKL1} \cite{BFKL2} \cite{BFKL3} is able to
describe very well the "hard Pomeron" (namely, the Pomeron in the region in
which the Mandelstam variables $s$ and $t$ are large enough to allow to
neglect non-perturbative effects. However, even this approach (which should be
considered as the first principle approach derived from QCD) fails when $s$
and $t$ are low enough (but still within the Regge region $\left\vert
s/t\right\vert \gg1$).

In the presence of a monopole, the background integral in the Regge
formula\footnote{As it is explained in the next section, this result together
with the cut induced by the monopole on the real axis of the complex $l$-plane
hold in the relativistic case as well.} in Eq. (\ref{SomWat2}) will not
decrease with $s$. Therefore, a clear-cut Regge pole behavior can only emerge
when the leading trajectory begins to dominate the background integral in Eq.
(\ref{SomWat2}). Consequently, the present results strongly suggest a proposal
which will be discussed in the next section.

\section{Relativistic generalizations}

It is natural to wonder whether the present results are just a curiosity of
the quantum mechanical setting considered in this paper or if they resist in a
QFT context as well. One can argue as follows that the second possibility is
likely to happen. When one considers the Froissart-Gribov extension of the
Regge formula in the QFT scattering of scalar particles in \textit{the
presence of a monopole} (or any other relevant topological defect such as
instantons, dyons and so on) very similar changes appear due to the fact that
the monopole modifies the generators of the angular momentum operator in the
QFT case too (see, for instance, \cite{sphericalupto1} \cite{Boulware}).
Indeed, according to the Gribov-Froissart procedure \cite{Gribov}
\cite{froissartP} in the scalar case when the presence of a monopole is taken
into account, it is natural to guess the following expression (following
chapter 1 of \cite{pomeronbook2}):%
\begin{align}
\left.  \mathit{A}_{\mu}\left(  s,t\right)  \right\vert _{\mu=1/2}  &
=2P_{l=0}^{1,0}\left(  1+2\frac{s}{t}\right)  a\left(  \lambda\left(
0\right)  ,t\right)  +\label{rel1}\\
&  \frac{1}{4i}%
{\displaystyle\int\limits_{C}}
\frac{\left[  2\left(  l+\mu\right)  +1\right]  dl}{\sin\pi l}\sum_{\eta=\pm
1}\left(  \eta+\exp\left(  -i\pi l\right)  \right)  P^{1,0}\left(
l,1+2\frac{s}{t}\right)  a^{\eta}\left(  \lambda\left(  l\right)  ,t\right)
\nonumber
\end{align}
where $P^{1,0}\left(  l,x\right)  $ is the analytical continuation in the
complex $l$-plane of the Jacobi polynomial $P_{l}^{1,0}\left(  x\right)  $ of
indices ($1$, $0$) and angular label $l$ while $a^{\pm1}\left(  \lambda\left(
l\right)  ,t\right)  $ are the analytic continuation of the even and odd
partial waves amplitudes and $\eta$ is the signature\footnote{It is worth to
note that, thanks to Eq. (\ref{usefulsignature}), the analysis of the behavior
of the partial wave amplitudes along the immaginary $l$-axis when $\left\vert
l\right\vert \rightarrow\infty$ is the same as in the usual case (see chapter
1 of \cite{pomeronbook2}).}. As in the non-relativistic case, the first term
cannot be included into the Sommerfeld-Watson contour $C$ due to the cut
related to the fact that the partial scattering amplitude $a\left(
\lambda\left(  l\right)  ,t\right)  $ depends on $l$ through $\lambda(l)$ in
Eq. (\ref{radialM0.5}). Thus, the Regge formula in this case reads%
\begin{align}
\left.  \mathit{A}_{\mu}\left(  s,t\right)  \right\vert _{\mu=1/2}  &
=2P_{l=0}^{1,0}\left(  1+2\frac{s}{t}\right)  a\left(  \lambda\left(
0\right)  ,t\right)  +\sum_{\eta=\pm1}\sum_{n_{\eta}}c_{n_{\eta}}%
P^{1,0}\left(  \alpha_{n_{\eta}}\left(  t\right)  ,1+2\frac{s}{t}\right)
\label{rel2}\\
+  &  \frac{1}{4i}%
{\displaystyle\int\limits_{\varepsilon-i\infty}^{\varepsilon+i\infty}}
\frac{\left[  2\left(  l+\mu\right)  +1\right]  dl}{\sin\pi l}\sum_{\eta=\pm
1}\left(  \eta+\exp\left(  -i\pi l\right)  \right)  P^{1,0}\left(
l,1+2\frac{s}{t}\right)  a^{\eta}\left(  \lambda\left(  l\right)  ,t\right)
\ ,\nonumber
\end{align}
where $\alpha_{n_{\eta}}\left(  t\right)  $ is the position of the $n_{\eta}%
$-th Regge pole of parity $\eta$ and all the coefficients multiplying the
Jacobi polynomials evaluated at the Regge poles (compare, for instance, with
Eq. (1.15) of \cite{pomeronbook2}) have been denoted as $c_{n_{\eta}}$ (which
depend on $\eta$ and on the transferred momentum $t$). Also in the
relativistic case the background integral in the Regge formula is very similar
to the one in Eq. (\ref{SomWat2}). Since the real part of $l$ in the last term
(namely, the background integral) of in Eq. (\ref{rel2}) is positive and
arbitrarily small while the imaginary part can vary from $-$ to $+\infty$, it
does not decrease when $s\rightarrow+\infty$ and a clear-cut Regge pole
behavior only emerges for large enough $t$ (assuming rising trajectories).

As from the theoretical point of view the main challenge is to explain the
observed behavior of the Pomeron for very low/vanishing $t$ and not too high
$s$ (namely, $s$ of the order of GeV or less), the concrete proposal arising
from the present results (and, in particular, Eq. (\ref{rel2})) is that
instead of parametrizing the soft Pomeron in terms of a single Regge pole (as
it is done in the majority of the phenomenological papers on this topic), one
should use a parametrization of the form in Eq. (\ref{rel2}) (keeping only the
leading Regge pole and the background integrals) to try to fit the
proton-proton scattering data in \cite{refinedPomeron}. The only missing piece
of information\footnote{It seems that there are not enough available data to
fully reconstruct $a\left(  \lambda\left(  l\right)  ,0\right)  $ yet.
However, the task to build a good phenomenological expression for $a\left(
\lambda\left(  l\right)  ,0\right)  $ does not appear to be out of reach.} is
the (re)construction of (a good phenomenological expression for) the partial
scattering amplitude $a\left(  \lambda\left(  l\right)  ,t\right)  $ for
vanishing $t$. Once this is achieved, it could be possible to reconcile the
observed non-decreasing scattering amplitude at $t=0$ (thanks to the
non-decreasing background integral) with a Pomeron trajectory with intercept
less than $1$ in agreement with unitarity (as, at least in the Schrodinger
case, the optical theorem with a monopole holds). This very interesting topic
is actually under investigation. It is worth to emphasize that any clear
deviation from the single-pole behavior at $t=0$ would make this proposal more attractive.

From a fundamental QCD perspective, the BFKL equation (introduced in
\cite{BFKL1} \cite{BFKL2} \cite{BFKL3}) is the best attempt to describe the
Pomeron from first principle (detailed review are \cite{pomeronbook2}%
\ \cite{LipaBook}). It describes the (Reggeized version of) the exchange of
two gluons contracted in such a way to have the vacuum quantum number as the
observed Pomeron. One of the basic building block of the BFKL equation is the
gluon propagator. Since, in the usual gluon propagator adopted in the BFKL
formalism, the internal and space-time indices \textit{are not linked} in
hedgehog-like style, it is impossible to incorporate into the BFKL formalism
this phenomenon of \textit{spherical symmetry up to internal rotations} (which
is one of the characteristic fingerprints of topologically non-trivial solitons).

The most obvious way to overcome this problem is to use in the BFKL equation
the gluon propagator obtained acting on the usual propagator with the singular
gauge transformations\footnote{The price to pay of course is that such gauge
transformations are singular and the corresponding singularities describe
magnetic monopoles degrees of freedom.} introduced in\ \cite{thooftsing}\ in
order to make the non-Abelian theory as "Abelian as possible" (the regularized
version of this gauge choice is now known as maximal Abelian gauge). Such
propagator (which would have the necessary hedgehog-like structure) would
include into the BFKL analysis relevant topological informations. Another
possibility would be to use in the BFKL equation the gluon propagator in the
background of an instanton or a non-Abelian monopole: the gluon propagator in
these cases would have the required hedgehog structure.

I hope to come back on this interesting issues in a future publications.

\section{Discussion and future developments}

\qquad In this paper the quantum mechanical scattering from a monopole has
been discussed using the Regge theory of complex angular momenta. In order to
apply the Regge theory in the presence of monopoles, a short range potential
(chosen as a superposition of Yukawa potentials) has to be included. Such an
inclusion is very welcome as it describes the typical effects of strong
interactions. Moreover, it is also useful as a description of the far field
behavior of the Higgs field of a 't Hooft-Polyakov monopole.

The results of the analysis is that the scattering amplitudes in the presence
of both a monopole and a short range potential are very similar to Jacob-Wick
helicity amplitude. From the intuitive point of view, this formal result is
quite satisfactory due to the well known fact \cite{sphericalupto1} that
scalars within a monopole field can behave as Fermions. However, the
application of Regge analytic continuation in the complex angular momentum
plane discloses a crucial difference between this case and the usual
Reggeization of helicity amplitude. Namely, a fixed branch cut on the real $l$
axis of width $2\left\vert \mu\right\vert $ (where $\mu$ is the strength of
the monopole) opens up. Such a cut is related to the modified generators of
the angular momentum operator in the presence of a topological soliton (and a
monopole in particular). The most relevant consequence of this fact is that
the background integral in the Regge formula cannot be pushed to the left.
Therefore, unlike what happens in the usual case, the background integral in
the Regge formula does not decrease anymore for large values of $\left\vert
\cos\theta\right\vert $. Hence, a clear-cut Regge pole behavior only emerges
when the transferred momentum is large enough so that the leading pole
dominates on the background integral. Consequently, at low transferred
momentum, the background integral plays a key role.

These results open the possibility to reconcile the observed non-decreasing
scattering amplitude at $t=0$ (thanks to the non-decreasing background
integral) with a Pomeron trajectory satisfying the unitarity constraint since,
at least in the Schrodinger case, the optical theorem with a monopole holds.
This very relevant topic is worth to be further investigated.

A further interesting issue is the analysis of the BFKL equation in the
presence of non-trivial topological structures such as monopoles. The present
results strongly suggest that the physical consequences of such topological
solitons are especially relevant at low transferred momentum, well within the
range of energies characterizing the soft Pomeron. A concrete way to include
such effects into the BFKL equation has been suggested.

At last, as it has been already remarked in the introduction, topological
defects in 2+1 dimensions give rise to similar effects as far as the
centrifugal barrier is concerned (a typical case being the interaction of a
charge with the (2+1)-dimensional magnetic monopole described by a
Chern-Simons term: see, for instance, the detailed review \cite{wilczek}).
Correspondingly, it is a very interesting and (to the best of author`s
knowledge) open question to extend the Regge formalism of complex angular
momenta to scattering problems by defects and short-range potential in (2+1)
dimensions. In particular, it would be very interesting to apply the Regge
formalism to the framework of \cite{plyus} in the presence of short range potentials.

\section{Acknowledgments}

\qquad The author wish to thank David Dudal, Vincent Mathieu, Pablo Pais,
Luigi Rosa, Cedric Troessaert and the anonymous referee for very useful
comments which improved this manuscript. This work has been funded by the
Fondecyt Grants No. 1160137. The Centro de Estudios Cient\i ficos (CECs) is
funded by the Chilean Government through the Centers of Excellence Base
Financing Program of Conicyt.

\end{document}